\documentclass[10pt]{article}
\include{epsf}
\def\be{\begin{equation}}
\def\ee{\end{equation}}
\def\bea{\begin{eqnarray}}
\def\eea{\end{eqnarray}}
\def\bnd{\partial X}
\def\bndl{|\partial X|}
\def\avbndl{\overline{|\partial X|}}
\def\vol{|\Lambda|}
\def\e0{|E_0|}
\def\Heff{H_{\rm eff}}
\begin{document}
\title{\large\bf Non-uniform ground state for the Bose gas}
\author{Andr\'as S\"ut\H o\\
Research Institute for Solid State Physics\\ Hungarian Academy of
Sciences \\ P. O. Box 49, H-1525 Budapest 114 \\ Hungary\\
E-mail: suto@szfki.hu}
\date{}
\maketitle
\thispagestyle{empty}
\begin{abstract}

\noindent

We study the ground state $\Psi_0=\sum a_X |X\rangle$
of $N$ hard-core bosons on a finite
lattice in configuration space, $X=\{x_1,\ldots,x_N\}$.
All $a_X$ being positive, the ratios
$a_X/\sum a_Y$ can be interpreted as probabilities $P_a(X)$.
Let $E_0$ denote the energy of the ground state and $|\partial X|$
the number of nearest-neighbour particle-hole pairs in the configuration $X$.
We prove the concentration of $P_a$ onto $X$ with $|\partial X|$ in a
$\sqrt{|E_0|}$-neighbourhood of $|E_0|$, show that the
average of $a_X$ over configurations with $\bndl=n$ increases exponentially
with $n$, discuss fluctuations about this average, derive upper and lower
bounds on $E_0$ and give an argument for
off-diagonal long-range order in the ground state.
\vspace{5mm}\\
PACS: 0530J, 0550, 7510J
\vspace{1cm}\\
Appeared in {\it J. Phys. A: Math. Gen.} {\bf 34} (2001) 37-55, {\it ibid.} {\bf 34} (2001) 6209
\end{abstract}
\newpage

\section{Introduction}

Bosons on a lattice interacting via an infinite on-site repulsion
(hard-core bosons) represent a system of double interest.
This is the simplest example of an interacting Bose gas and, thus,
the most promising candidate for a rigorous treatment of Bose-Einstein
condensation of interacting particles.
The model is also known to be equivalent to a
system of ${1\over 2}$ spins 
(Matsubara and Matsuda 1956) coupled via the $X$ and $Y$ and
possibly the $Z$ components of neighbouring spins and exposed to an external
magnetic field in the $Z$ direction. Ordering of the planar component of
the spins is equivalent to Bose-Einstein condensation or the appearance of
an off-diagonal long-range order (ODLRO, Yang 1962) in the system of bosons.
Apart from some exceptions, such as bounds on the density of the condensate
(T\'oth 1991) or the discussion of the model on complete graphs
(T\'oth 1990, Penrose 1991),
the most interesting rigorous results were formulated
in spin terminology and obtained by the use of a particular
symmetry, reflexion positivity (Dyson {\it et al} 1978, Kennedy {\it et al} 1988,
Kubo and Kishi 1988).
This introduced limitations on
the value of the external field (zero field) and the lattice type (essentially
hypercubic lattices). Expressed in terms of a boson gas, ODLRO
was shown only at half filling on hypercubic lattices: in the ground state
in dimensions $\geq 2$, and for positive temperatures above two
dimensions. The proof of ordering does not offer much insight into the
structure of the state. The only case in which details are known is the
mathematically nice but physically trivial example of the complete graph
(T\'oth 1990, Penrose 1991).

In this paper we provide informations concerning the ground state $\Psi_0$,
valid for very different lattices and arbitrary particle
densities. The main result is the proof of a strong non-uniformity of
$\Psi_0$ in configuration space. It appears mathematically in the
form of a large-deviation principle and makes the ground state
resemble a thermal Gibbs distribution of a classical gas on the same lattice.
It also leads to an approximate expression for $\Psi_0$ and to
an argument for Bose-Einstein
condensation in the ground state in two and higher dimensions.

Let $\bf L$ be an infinite
lattice which, for the sake of simplicity, will be supposed to be regular
with a constant coordination number (degree) $k$. Throughout the paper
$\Lambda$ denotes a finite connected part of $\bf L$
taken with periodic boundary
conditions so as to keep the degree constant (not really essential). The
Hamiltonian we are going to study in detail is
\begin{equation}\label{Ham}
H_0=-\sum_{\langle xy\rangle}(b^*_x b_y+b^*_y b_x)\ .
\end{equation}
We write $x,y,\ldots$ for the vertices (sites) and $\langle xy\rangle$ for the
edges (nearest-neighbour pairs); $b^*_x$ and $b_x$ create, respectively, annihilate a
hard-core boson at $x$. Boson operators at different sites commute with
each other while
\begin{equation}\label{hc}
b_x^*b_x+b_xb_x^*=1
\end{equation}
accounts for the hard-core condition.
Correspondence with spin models is obtained
by setting $b_x=S_x^-$ and $b_x^*=S_x^+$. The Hamiltonian
conserves the number of bosons,
\begin{equation}\label{N}
N=\sum_{x\in\Lambda}n_x=\sum_{x\in\Lambda}b_x^*b_x
\end{equation}
and is also invariant under particle-hole transformation. We can, therefore,
fix $N$ so that $\rho=N/|\Lambda|$ is between 0 and ${1\over 2}$. (Here
and below, if $A$
is a finite set, $|A|$ stands for the number of elements.)
Let $X,Y,\ldots$ denote $N$-point
subsets of $\Lambda$, also called configurations.
A convenient basis is formed by the states
\begin{equation}\label{basis}
|X\rangle=\left(\prod_{x\in X}b_x^*\right)|0\rangle
\end{equation}
where $|0\rangle$ is the vacuum state.
According to the Perron-Frobenius theorem, there is a single ground state
$\Psi_0$ whose
all coefficients in the basis (\ref{basis}) are positive. Let $a=(a_X)$ denote
the vector of these coefficients. Since the unique source of energy is
hopping constrained by an on-site exclusion, we can expect
that $a_X$ increases, at least `on average', with an increasing
number of nearest neighbour particle-hole pairs.
More precisely,
let $\partial X$ denote the set of nearest-neighbour particle-hole pairs
for $X$ (the `boundary' of $X$) and $\Omega_n$ the ensemble
of configurations $X$ with `boundary length' $\bndl=n$.
It is easily seen that
\be\label{bndl}
\bndl=\langle X|H_0^2|X\rangle=\langle X|\Delta|X\rangle
\ee
where
\be\label{diag}
\Delta=\sum_{\langle xy\rangle}[n_x(1-n_y)+n_y(1-n_x)]
\ee
is the diagonal part of $H_0^2$, and $|\Omega_n|$ is the multiplicity of the
eigenvalue $n$ of $\Delta$.
We expect that the $X$-dependence of $a_X$ appears mainly through $\bndl$,
and the average
\be\label{alfa}
\alpha_n={1\over|\Omega_n|}\sum_{X\in\Omega_n}a_X
\ee
is some increasing function of $n$.
We shall indeed find that $\alpha_n$ rapidly grows with $n$ -- exponentially
fast for $n$ of the order of $N$ and even faster for smaller $n$.
To make a comparison with the free Bose gas, let us note that the ground state
of the latter reads
\be
\Psi_{\rm free}=\vol^{-N/2}\sum_{\{m_x\geq 0\}_{x\in\Lambda}:\sum m_x=N}
\sqrt{N!\over\prod_{x\in\Lambda}m_x!}\prod_{x\in\Lambda}(a_x^*)^{m_x}|0\rangle
\ee
where $a_x^*$ is the ordinary boson creation operator at $x$. If we project
out the part with no particle encounters, we find
constant $\times\sum|X\rangle$, the sum going over the basis (\ref{basis}).
This uniform sum is qualitatively very different from what we actually find for the
ground state, and the energy of this state will also be found to be extensively
higher than that of the ground state.

We start, in section 2, by studying the probability measure
\be\label{pan}
p_a(n)={|\Omega_n|\alpha_n\over\sum_m|\Omega_m|\alpha_m}\ .
\ee
We show that the mean value according to $p_a$ is $|E_0|$, the modulus of
the ground state energy, while
the mean square deviation $D_a^2$ is of the order of $N$, so that $p_a(n)$
is concentrated onto integers with $|n-|E_0||$ of the order of $\sqrt{N}$.
In section~3 we evoke a large deviation principle for the Ising
model on the same lattice, having the approximate form
\be\label{qan}
q(n)\equiv{|\Omega_n|\over \sum_m|\Omega_m|}\propto e^{-{(n-M)^2\over 2D^2}}\ .
\ee
Here $M=\avbndl$, the (arithmetic)
mean of boundary lengths among $N$-point configurations
(and also the modulus of the energy of the projected free-boson
ground state) and $D^2$ is the
corresponding mean square deviation of $\bndl$.
We derive a formula for $D^2$ as a function of $\rho$ and $k$
($M=k\rho(1-\rho)|\Lambda|$ can be obtained trivially).
In section~4 we use the results on $p_a$ and $q$ to
give an approximate expression
for $\alpha_n$ in terms of $E_0$, $D_a$, $M$ and $D$. We present an
argument for the monotonicity of $\alpha_n$ and deduce from it
upper and lower bounds on
$D_a/D$, predicting $\alpha_n=\ $constant if $|E_0|/M=1$
asymptotically (as $\vol$ and $N$ go to infinity).
The variational treatment
in section~5 shows that $|E_0|/M>1$ in the thermodynamic limit.
A notable exception, with $|E_0|=M=N(|\Lambda|-N)$,
is the complete graph where by
permutational symmetry we actually have $a_X=\ $constant. In section~5
we also present lower bounds on $E_0$ for all densities. The physical
consequences of our findings are resumed in section~6.
Here we provide an argument for the existence of ODLRO in the
two and higher dimensional models. It is based on a hypothesis which also
permits to estimate the dimension dependence of the deviations of
$a_X$ from the average $\alpha_{\bndl}$ and to conclude that these deviations
are irrelevant above two dimensions.
In section~7 we briefly discuss extensions to
Hamiltonians with more complicated interactions.

\section{Non-uniformity of the ground state}

The results of this section are valid for any connected, not necessarily
regular lattice with all coordination numbers $\leq k$.

Let $A=(A_{XY})$ denote the matrix of $-H_0$ in the basis (\ref{basis}).
Then $A$ is the adjacency matrix of a graph $G$
whose vertices are the configurations, and two configurations $X$ and $Y$ form
an edge of $G$
if and only if their symmetric difference is an edge of $\Lambda$:
$X\setminus Y\cup Y\setminus X=\{x,y\}=\langle xy\rangle$.
We denote the largest eigenvalue of $A$ by
$\lambda_1$ and the corresponding eigenvector by $a=(a_X)$.
The ground state energy and wave function are related to them through
$E_0=-\lambda_1$ and $\Psi_0=\sum a_X|X\rangle$.
The coefficients of the ground state wave vector $a$ being positive,
$a_X/\sum a_Y$ can be interpreted as a probability $P_a(X)$ of $X$.
As we shall see, $P_a(X)$ is very far from being a uniform distribution
that one can find on complete graphs.

A well-known fact concerning adjacency matrices is that $(A^n)_{XY}$ provides
the number of walks of length $n$ in $G$ between $X$ and $Y$ (Biggs 1974).
Let
\be\label{W}
W_n(X)=\sum_Y(A^n)_{XY}\ ,
\ee
the number of walks of length $n$ starting from $X$. First, we note that
the expectation value of this number with respect to $P_a$,
\be\label{EaWn}
\langle W_n\rangle_a\equiv \sum_X W_n(X)a_X/\sum_X a_X=\lambda_1^n
=\langle W_1\rangle_a^n\ .
\ee
This follows by taking the $X$ component of the vector equation
$A^n a=\lambda_1^n a$, summing over $X$ and dividing by $\sum a_X$.
The above equality holds true for any simple graph: the probability measure
$P_a$ `sees' the graph as if it were regular with a degree $\lambda_1$.

Next, we make use of the fact that actually we are dealing with a sequence
of graphs, $G=G_{\Lambda,N}$, having the particular property that the
typical degree is of order $N$ while its change between
neighbouring vertices is of order one:
The degree of $X$ is its number of neighbours,
$W_1(X)$. There is a one-to-one correspondence between nearest neighbour
particle-hole pairs if $X$ is considered as a subset of $\Lambda$
and neighbours of $X$ as a vertex in $G$. Therefore, $\bnd$
can be identified with the set of neighbours of $X$ in $G$ and we have
$W_1(X)=\bndl$.
When passing from $X$ to a $Y\in\bnd$ a neighbouring particle-hole pair
is interchanged. For both the particle and the hole
the number of neighbours of the opposite kind can change by
at most $k-1$, whence
\be\label{parxy}
|W_1(X)-W_1(Y)|\leq 2(k-1) \quad\mbox{if}\quad Y\in\partial X\ .
\ee
Now
\be\label{W2}
W_2(X)=\sum_{Y\in\partial X}|\partial Y|
\ee
which together with (\ref{parxy}) yields
\be\label{w1s-w2}
|W_1(X)^2-W_2(X)|\leq 2(k-1)\bndl\ .
\ee
Taking the expectation value and using (\ref{EaWn}),
\be\label{Da}
0\leq D_a^2\equiv \langle W_1^2\rangle_a-\langle W_1\rangle_a^2
\leq 2(k-1)\lambda_1\ .
\ee
By Chebyshev's inequality we then find that for any $\varepsilon>0$
\be\label{Cheb}
P_a\left(|W_1-\lambda_1|>\sqrt{2(k-1)\lambda_1/\varepsilon}\right)<\varepsilon
\ee
or equivalently
\be\label{Ch2}
\sum_{|W_1(X)-\lambda_1|\leq\sqrt{2(k-1)\lambda_1/\varepsilon}}a_X
\geq(1-\varepsilon)\sum a_X\ .
\ee
Similarly to (\ref{W2}),
\be\label{Wn}
W_n(X)=\sum_{Y_1\in\partial X}\sum_{Y_2\in\partial Y_1}\ldots
\sum_{Y_{n-1}\in\partial Y_{n-2}}|\partial Y_{n-1}|
\ee
which, together with (\ref{parxy}), yields
\be\label{Wnineq}
W_n<\prod_{l=0}^{n-1}(W_1+2(k-1)l)\qquad W_n>\prod_{l=0}^{n-1}(W_1-2(k-1)l)\ .
\ee
These bounds are nontrivial if $n\ll |\Lambda|$, for example, for $n$ fixed and
$|\Lambda|$ going to infinity. (In the opposite limit we have the stronger
inequality $W_n(X)\leq (kN)^n$.)
From (\ref{Wnineq}) we get
\be\label{w1n-wn}
|W_1^n-W_n|
\leq\sum_{m=1}^{n-1}(2k-2)^m
\left(\sum_{1\leq l_1<\cdots<l_m\leq n-1}l_1\cdots l_m\right) W_1^{n-m}\ .
\ee
Taking the expectation value,
\be
|\langle W_1^n\rangle_a-\lambda_1^n|
\leq\sum_{m=1}^{n-1}(2k-2)^m
\left(\sum_{1\leq l_1<\cdots<l_m\leq n-1}l_1\cdots l_m\right)
\langle W_1^{n-m}\rangle_a\ .
\ee
Replacing $n$ by $n-m$ we obtain
\be
\langle W_1^{n-m}\rangle_a\leq \lambda_1^{n-m}+O(\lambda_1^{n-m-1})\ .
\ee
Now $\lambda_1$ is of order $|\Lambda|$, so finally
\be\label{Ean}
|\langle W_1^n\rangle_a-\lambda_1^n|\leq
 n(n-1)(k-1)[1+O(|\Lambda|^{-1})] \lambda_1^{n-1}\ .
\ee
Equations~(\ref{Da})-(\ref{Ch2}) and (\ref{Ean})
describe the concentration of the probability measure $P_a$ on configurations
$X$ whose boundary length $|\partial X|$ is in a $\sqrt{N}$-neighbour\-hood
of $|E_0|$. We shall refer to this property as the non-uniformity of the
ground state.
Summing over $X$ in $\Omega_n$, $P_a$ gives rise to the
probability distribution $p_a$, which is, hence, peaked about $|E_0|$.
In section 4 we shall arrive at the same conclusion in a different way,
by studying first
the {\em a priori} distribution $q(n)=|\Omega_n|/\sum_m|\Omega_m|$.

\section{Large-deviation principle for the a priori distribution}

The precise computation of $|\Omega_n|$ is a difficult and unsolved
combinatorial problem. The logarithm of this number is
the entropy of the Ising model, for energy $n$,
in a microcanonical ensemble with a fixed
magnetization, $\sum_{x\in\Lambda}\sigma_x\!=2N-|\Lambda|$.
Indeed, if we put $\sigma_x=1$ for $x$ in $X$ and $\sigma_x=-1$ elsewhere,
we get an Ising configuration with a density $\rho$ of + spins.
The corresponding union of contours can be identified with
$\partial X$ whose total length $|\partial X|$ is the
energy of the Ising configuration. For such systems there exists a strong
version of the equivalence of ensembles which, applied to our case, states
that the distribution $q$ of $|\partial X|$
satisfies a large deviation principle
whose rate function is, apart from a shift, minus
the specific entropy $s(e,\rho)$
of the Ising model ($e=n/\vol$), cf. Pfister 1991, Dobrushin and
Shlosman 1994 and Lewis {\it et al} 1994.
The exact form of the entropy is unknown. To circumvent this problem,
we use an approximate formula for the probability of having a boundary
of length $n$,
\begin{equation}\label{ldp}
q(n)\approx
Z^{-1}\exp\{-{(n-M)^2\over 2D^2}\}\ .
\end{equation}
Here $Z\propto\sqrt{N}$ is for normalization and
\begin{equation}\label{MD2}
M=\overline{|\partial X|}=k\rho(1-\rho)|\Lambda|\qquad
D^2=\overline{(|\partial X|-\overline{|\partial X|})^2}\ ,
\end{equation}
cf. Eq.~(\ref{qan}) in the Introduction.
The Gaussian approximation is correct in a neighbourhood of the
maximum of the specific entropy if
$|\Lambda|/D^2$
is non-vanishing in the thermodynamic limit.
This is what we are going to check by explicitly computing
the mean square deviation of $\bndl$. We show that
\begin{equation}\label{D2}
D^2=[k-2(2k-1)\rho(1-\rho)]k\rho(1-\rho)|\Lambda|=
[k-2(2k-1)\rho(1-\rho)]M\ .
\end{equation}
We have found this expression first for $d$-dimensional hypercubic lattices,
and have also checked it for the triangular, honeycomb and Kagom\'e
lattices. The apparent independence of the details
of the lattice is somewhat surprising because
our derivation below needs knowledge of the rather different local
neighbourhoods up to next-nearest neighbours. The only common feature of all
these lattices seems to be that all sites are symmetry-related and thus equivalent.
This alone should therefore suffice to prove Eq.~(\ref{D2}).
Let us note that, with the microcanonical temperature defined as
$(\partial s/\partial e)^{-1}$, the maximum of the entropy corresponds to
infinite temperature. Therefore, in a neighbourhood of the maximum we are
at a safe distance to the ferro- and antiferromagnetic phase transitions,
which could weaken our approximation through a breakdown (in finite volumes)
of the concavity of the microcanonical entropy (Pleimling and H\"uller 2000).

Equation (\ref{D2}) is obtained by filling the sites of $\Lambda$ independently
and with equal probability $\rho$. We expect smaller order corrections to
appear if the computation is done with $N$ fixed, $N/\vol=\rho$ (cf. Eq.~(\ref{avbound}) below).
So in this section $X$ is a random subset of $\Lambda$ whose probability
to be selected is $\rho^{|X|}(1-\rho)^{|\Lambda|-|X|}$,
and $n_x=n_x(X)$ is a random variable taking the value 1 if $x$ is in $X$
and 0 otherwise. Then all $n_x$ are independent
and take 1 with probability $\rho$ and 0 with probability $1-\rho$.

We define
\begin{equation}\label{fx}
f_x=n_x\sum_{y\in\partial x}(1-n_y)
\end{equation}
where $\partial x$ denotes the set of neighbours of $x$ in $\Lambda$.
The boundary length of $X$ is obtained (cf. Eqs.~(\ref{bndl}) and (\ref{diag})) as
\begin{equation}\label{DeltaX}
|\partial X|=\sum_{x\in \Lambda}f_x(X)\ .
\end{equation}
Thus the mean value of (\ref{DeltaX}) is
\begin{equation}\label{M}
M=\sum_{x\in\Lambda}\overline{f_x}=k\rho(1-\rho)|\Lambda|
\end{equation}
as claimed earlier.

Let $d(x,y)$ denote the graph distance of $x$ and $y$ in $\Lambda$, i.e.
the length of the shortest walk between them. Since $f_x$ and $f_y$ are
independent if $d(x,y)>2$, we find
\begin{equation}\label{mean}
D^2=\sum_{x,y}\overline{(f_x-\overline{f_x})(f_y-\overline{f_y})}=
\sum_{x,y:\,d(x,y)\leq 2}r(x,y)
\end{equation}
\begin{equation}\label{cov}
r(x,y)=\overline{f_xf_y}-\overline{f_x}^2\ .
\end {equation}

The computation of the different terms is straightforward by observing
that $n_x^2=n_x$.
The contribution of the diagonal terms $x=y$ is the same for any $k$-regular
lattice. Namely,
\begin{equation}\label{d=0}
\overline{f_x^2}=k\rho(1-\rho)+k(k-1)\rho(1-\rho)^2
\end{equation}
\begin{equation}\label{c0}
\sum_{x\in\Lambda}r(x,x)=|\Lambda|r(x,x)=
[k-(2k-1)\rho+k\rho^2]M\ .
\end{equation}
The contribution of nearest neighbour pairs depends on the number of triangles
containing a given edge. If there are $\ell$ such triangles then
\begin{equation}\label{d=1}
\overline{f_xf_y}=\rho^2[\ell(1-\rho)+[(k-1)^2-\ell\,](1-\rho)^2]
\end{equation}
\begin{equation}\label{c1}
\sum_{x,y:\,d(x,y)=1}r(x,y)=k|\Lambda|r(x,y)
=\rho[\ell\rho-(2k-1)(1-\rho)]M\ .
\end{equation}
If $x$ and $y$ are next-nearest neighbours to each other,
they may have $m$ common nearest neighbours. Then
\begin{eqnarray}\label{d=2}
\overline{f_xf_y}&=&\rho^2[m(1-\rho)+(k^2-m)(1-\rho)^2]\nonumber\\
r(x,y)&=&m\rho^3(1-\rho)\ .
\end{eqnarray}
In $d$-dimensional hypercubic lattices ($k=2d$) there are next-nearest
neighbour pairs with $m=1$ and $m=2$. Their contribution to $D^2$ is
\begin{eqnarray}\label{c2Zd}
\sum_{x,y:\,d(x,y)=2}r(x,y)&=&k|\Lambda|\,r(x,y)_{m=1}+4{d\choose2}|\Lambda|\,
r(x,y)_{m=2}\nonumber\\
&=&\rho^2 M+(k-2)\rho^2 M=(k-1)\rho^2 M\ .
\end{eqnarray}
For the triangular lattice ($k=6$)
\begin{equation}\label{c2tr}
\sum_{x,y:\,d(x,y)=2}r(x,y)=k|\Lambda| [r(x,y)_{m=1}+r(x,y)_{m=2}]
=(k-3)\rho^2 M\ .
\end{equation}
In the honeycomb lattice ($k=3$) each site has six next-nearest neighbours,
all of the type $m=1$. So
\begin{equation}\label{c2hon}
\sum_{x,y:\,d(x,y)=2}r(x,y)=2k|\Lambda|\,r(x,y)_{m=1}=(k-1)\rho^2 M\ .
\end{equation}
In the Kagom\'e lattice ($k=4$) there are eight next-nearest neighbours with
$m=1$:
\begin{equation}\label{c2Kag}
\sum_{x,y:\,d(x,y)=2}r(x,y)=2k|\Lambda|\,r(x,y)_{m=1}=(k-2)\rho^2 M\ .
\end{equation}

Finally, we obtain $D^2$ by adding (\ref{c0}) and (\ref{c1}) with $\ell=0$
and (\ref{c2Zd}) for hypercubic lattices,
(\ref{c0}) and (\ref{c1}) with $\ell=2$ and (\ref{c2tr})
for the triangular lattice, (\ref{c0}) and (\ref{c1}) with $\ell=0$ and
(\ref{c2hon}) for the honeycomb lattice and (\ref{c0}), (\ref{c1}) with
$\ell=1$ and (\ref{c2Kag}) for the Kagom\'e lattice. All yield (\ref{D2}).

If the random variables $f_x$ were independent, the mean square deviation
of their sum would be given by (\ref{c0}). For any $\rho\leq{1\over 2}$
this is larger than the actual value (\ref{D2}), so on average the $f_x$
are negatively correlated. Because $\ell\leq k-1$, the nearest
neighbour correlation $r(x,y)<0$ for all $\rho\leq{1\over2}$, cf.
Eq.~(\ref{c1}). On the other hand, according to (\ref{d=2}), for
next-nearest neighbours $r(x,y)$ is always positive.
Apparently, the induced probability distribution for the $f_x$ could be
approximated by a Boltzmann-Gibbs distribution written with an
antiferromagnetic nearest-neighbour Hamiltonian for the centralized
variables $f_x-\overline{f_x}$ and with a temperature and an external field
chosen so as
to fit the computed average (\ref{M}) and nearest-neighbour correlations
(\ref{c1}). If the effective $\beta$ and field are below their
respective critical values we get exponentially decaying antiferromagnetic
correlations, and the
approximation is qualitatively correct. Adding
next-nearest-neighbour interactions one could fit the
computed next-nearest-neighbour correlations, and so on.

\section{The form of the averaged wave vector}

Using the large-deviation result for $q(n)$ we can obtain more precise
information on $p_a(n)$ and $\alpha_n$.
Apart from normalization $p_a(n)$ is obtained by multiplying $q(n)$ with
$\alpha_n$.
Since, as we shall see in the next section, $\lambda_1>M$
and the difference is of the order of $N$,
$\alpha_n$ has to increase exponentially fast -- at least in a neighbourhood
of $M$ -- so as to shift the expectation
value $M$ of $q$ to the expectation value $\lambda_1$ of $p_a$. As a
result, we obtain the approximate expression
$p_a(n)\sim\exp -(n-\lambda_1)^2/2D_a^2$, consistent with our findings in
section 2.
This is a second, independent, argument for the non-uniformity
of the ground state, assuming the form
of another large-deviation principle for $\bndl$.
To get $\alpha_n$ we equate this form of $p_a(n)$
with the other one, (\ref{pan}),
where $|\Omega_n|$ is estimated through (\ref{qan}). This gives
\be\label{alen}
\alpha_n\sim
\exp\left({(n-M)^2\over 2D^2}-{(n-\lambda_1)^2\over 2D_a^2}\right) \propto
       \exp{(D_a^2-D^2)n^2+2(D^2\lambda_1-D_a^2M)n\over 2D^2D_a^2}
\ee

For the moment, we have only trivial estimates for $\lambda_1$ ($\geq M$) and $D_a$
(cf. Eq.~(\ref{Da})). More precise bounds on $D_a$ can be obtained from the monotonic
increase of $\alpha_n$.

The argument telling that $\alpha_n$ should increase for all allowed $n$ is as follows.
From the spectral decomposition of $A^m$ we obtain (supposing $\sum a_X^2=1$)
\be\label{nonbipar}
a_X a_Y=\lim_{m\rightarrow\infty}\lambda_1^{-m}(A^m)_{XY}
\ee
if $\Lambda$ is non-bipartite, and
\be\label{bipar}
a_X a_Y=\lim_{m\rightarrow\infty}{1\over 2}[\lambda_1^{-m}(A^m)_{XY} +
                 \lambda_1^{-m-1}(A^{m+1})_{XY}]
       =\lim_{m\rightarrow\infty}{\lambda_1^{-r_m}\over 2}(A^{r_m})_{XY}
\ee
if $\Lambda$ is bipartite; $r_m=2m$ or $2m+1$ if the
graph distance $d_G(X,Y)$ of $X$ and $Y$ is even or odd, respectively.
Summing over $Y$ and averaging over $X$ in $\Omega_n$ we obtain, for example, for a
non-bipartite $\Lambda$,
\be
\alpha_n=(\sum a_Y)^{-1}\lim_{m\rightarrow\infty}
\lambda_1^{-m}{1\over |\Omega_n|}\sum_{X\in\Omega_n}W_m(X)\ .
\ee
Inspecting Eq.~(\ref{Wn}) one can conclude that
for any $m$ the average of $W_m(X)$ over $\Omega_n$ increases with $n$
and, hence, $\alpha_n\geq\alpha_{n'}$ if $n>n'$.

Let $n_{\rm min}$ and $n_{\rm max}$ denote the smallest and largest
allowed values
of $\bndl$, respectively. Since $n_{\rm min}=o(N)$, monotonicity of $\alpha_n$
implies through (\ref{alen})
\be\label{lowup}
{n_{\rm max}-\lambda_1\over n_{\rm max}-M}\leq{D_a^2\over D^2}\leq{\lambda_1\over M}
\ee
up to an error of $o(1)$. Therefore, the coefficient of $n$ in the exponent of $\alpha_n$
is non-negative.
Clearly $\lambda_1/M=1$ would imply $D_a/D=1$ and
$\alpha_n=\ $constant.
In the next section we show that for finite dimensional lattices $\lambda_1/M>1$
asymptotically and, hence, the two bounds in (\ref{lowup}) form an interval around~1.
On bipartite lattices $n_{\rm max}=kN$, so at half filling $n_{\rm max}=2M$ and
1 is in the centre of the interval. It is possible that $D_a=D$ and the exponent
of $\alpha_n$ is linear in $n$.
However, even if $D_a\neq D$, the quadratic term
in the exponent is of the same order as or even smaller (if $n=o(N)$) than
the linear term.

More information can be extracted from Eqs.~(\ref{nonbipar}) and (\ref{bipar})
if we combine them with (\ref{Ch2}). If $m$ is large enough,
\be\label{drift}
1-2\varepsilon\leq{1\over W_m(X)}\sum_{||\partial Y|-\lambda_1|\leq
\sqrt{2(k-1)\lambda_1/\varepsilon}}(A^m)_{XY}\leq 1\ ,
\ee
that is, for any $X$ and any sufficiently large $m$ an overwhelming
majority of walks of length $m$ starting from $X$
on the graph $G$ end up in a vertex whose degree is
$\lambda_1+O(\sqrt{N})$. In the expression
\be\label{w=a}
W_m(X)=(a_X \sum a_Y)\lambda_1^m+o(\lambda_1^m)\quad (m\rightarrow\infty)\ ,
\ee
obtained from (\ref{nonbipar}),
$\lambda_1^m$ accounts for this `long-time' behaviour.
Let us note that (\ref{drift}) cannot be understood
by imagining the walks on $G$ as realizations
of some simple random process. Because $G$ is nonregular (unless $\Lambda$
is a complete graph), no Markov process can assign equal probabilities to all
walks of equal length. As an example, for a locally unbiased random motion
(which chooses among neighbours with equal probability) the most probable
individual walks of a given length are the
`descending' ones, those going towards vertices of a lower degree, and
the less probable the `ascending' ones.
Because of the form (\ref{ldp}) of the distribution of
degrees, an $X$ with $\bndl<M$ has typically more ascending than
descending neighbours and {\em vice versa} for $\bndl>M$. Therefore,
the most probable degrees of end-points of very long walks would be
nevertheless close to $M$ (but not to $\lambda_1$).

Although the monotonicity of $\alpha_n$ holds for all $n$ between $n_{\rm min}$ and
$n_{\rm max}$, the validity of (\ref{alen}) is limited to a neighbourhood of the
interval $[M,\lambda_1]$.
In particular, the growth of $\alpha_n$ for
$n=o(N)$ is much steeper than the exponential predicted by (\ref{alen}). Indeed,
let us write the eigenvalue equation in the form
\be\label{eigen}
{1\over \bndl}\sum_{Y\in\bnd}a_Y={\lambda_1\over\bndl}a_X\ .
\ee
For a $d$-dimensional lattice $n_{\rm min}=O(N^{d-1\over d})$. If we choose
$\bndl=n$ to be of this order, we find that the average of $a_Y$ over $\bnd$ yields
$N^{1/d}$ times $a_X$! Since $q(n)$ is also rapidly increasing
here, the average over $\bnd$ is dominated by $Y$ with $n+2\leq |\partial Y|\leq n+2k-2$.
From this and the monotonicity of $\alpha_n$ we conclude that
$\alpha_{n+2k-2}/\alpha_n$ is at least of order $N^{1/d}$.

\section{Bounds on the ground state energy}

\subsection{Variational upper bounds}

Variational estimates of
the ground state energy are of the form
\begin{equation}\label{var}
E_0\leq\langle\psi|H|\psi\rangle/\langle\psi|\psi\rangle\ .
\end{equation}
A trivial choice is
\begin{equation}\label{triv}
\psi=\sum |X\rangle
\end{equation}
with the summation going over all $N$-point subsets of $\Lambda$. It yields
$E_0\leq -M$ where
\begin{equation}\label{avbound}
M=\avbndl={2|EG|\over |VG|}={2|E\Lambda|{\vol -2\choose N-1}\over
{\vol\choose N}}=k(1-\rho)N+O(1)\ ,
\end{equation}
cf. Eq.~(\ref{M}).
Here $|EG|$ and $|VG|$ denote the number of edges and vertices of $G$,
respectively, and $|E\Lambda|$ the number of edges of $\Lambda$.
It is not {\em a priori} obvious that this bound can be improved in the
order of the volume, and it is important to know that it really can.
In the opposite case, if $-k\rho(1-\rho)$ were
the true ground state energy per site then,
as in the complete graph, the Hamiltonian (\ref{Ham}) would have a
product ground state in an infinite volume,
\begin{equation}\label{prod}
\Psi=\prod_x (\sqrt{\rho}\ |n_x=1\rangle+\sqrt{1-\rho}\ |n_x=0\rangle)\ ,
\end {equation}
i.e. no local perturbation could decrease the energy
of $\Psi$. (Because of the product structure, if the energy could be
decreased locally, the specific energy $-k\rho(1-\rho)$ of $\Psi$
could also be decreased.) Since $\Psi$ shows
ODLRO with the value of the order parameter at its theoretical maximum
($\rho(1-\rho)$, cf. Appendix),
by proving it cannot be a ground state
we exclude a trivial scenario for Bose-Einstein condensation.

To prove that $|E_0|-M$ is of order $N$ we apply
trial functions of the form
\begin{equation}\label{trial}
\psi_v=\sum v_X|X\rangle \quad\qquad v_X=v(|\partial X|)
\end{equation}
i.e. $v_X$ depending on $|\partial X|$ only.
The variational bound (\ref{var}) reads $\lambda_1\geq B(v)$ where
\begin{equation}\label{Bv}
B(v)\equiv {(v,Av)\over(v,v)}
={\sum_n v(n)\sum_{i=-k+1}^{k-1}v(n+2i)\sum_{X\in\Omega_n}N_i(X)
\over\sum_n v(n)^2 |\Omega_n|}\ .
\end{equation}
Here $N_i(X)$ is the number
of those neighbours of $X$ having a boundary length $|\partial X|+2i$.
In Eq.~(\ref{Bv})
we have used (\ref{parxy}) and the fact that the parity of $\bndl$ is
the same for all $X$: even if $N$ is even and that of $k$ if $N$ is odd.

The form (\ref{alen}) of $\alpha_n$ suggests that the best choice for $v(n)$
would be $e^{xn+yn^2}$ with $x$ of order 1 and $y$ of order ${1\over N}$.
However, we do not expect the
quadratic term to yield a significant improvement, and choose a simple
exponential $v(n)=e^{xn}$ with $0<x<x_{\rm max}$ where
\begin{equation}\label{xmax}
x_{\rm max}={n_{\rm max}-M\over 2D^2}\ .
\end{equation}
Then by making use of the large-deviation principle for $\bndl$ we find
\be\label{av*av}
B(v)=\langle n\rangle_x \langle e^{2xi}\rangle_{[\langle n\rangle_x]}
\ee
where
\be
\langle n\rangle_x={\sum n e^{2xn} |\Omega_n|\over\sum e^{2xn} |\Omega_n|}\qquad
\langle f(i)\rangle_n=\sum_{i=-k+1}^{k-1}f(i){1\over n|\Omega_n|}\sum_{X\in\Omega_n}N_i(X)\ .
\ee
Now, since $\langle n\rangle_0=M$ and
$d\langle n\rangle_x/dx\left|_{x=0}=2D^2\right.$, we find
\be
\langle n\rangle_x=M(1+2xD^2/M+O(x^2))\ .
\ee
We note that the approximate form (\ref{qan}) would yield the same
result.

Next, we turn to the second term of $B(v)$. Obviously
$\langle e^{2xi}\rangle_n\geq e^{-2(k-1)x}$. This yields
$B(v)/M>1$ for $\rho$ near 0 but not near $1\over 2$. However, we can use
Jensen's inequality $\langle e^{2xi}\rangle_n\geq e^{2x\langle i\rangle_n}$
together with the fact that $\langle i\rangle_{[M]}$ goes to zero as
the volume increases. We then conclude that asymptotically
$\langle i\rangle_{[\langle n\rangle_x]}=-\ {\rm constant}\times x$ $+$ higher
order terms, so that the average of the exponential in Eq.~(\ref{av*av}) equals
$1-O(x^2)$. Thus, for $x$ small enough we indeed find $B(v)/M>1$.

With somewhat more effort one can actually compute a lower bound on $B(v)$.
We do not make it here, only notice that
in optimizing such a bound according to $x$ (and also in optimizing
a lower bound on $E_0$, see below)
the knowledge of $n_{\rm max}$ is necessary. On bipartite lattices
$n_{\rm max}=k\rho |\Lambda|$ for all $\rho\leq {1\over 2}$,
and it is an easy graphical exercise to see that the same
equality holds for $\rho\leq {1\over 3}$ on the triangular and
Kagom\'e lattices. However, for both lattices $n_{\rm max}$ is
constant between the
densities ${1\over 3}$ and ${1\over 2}$: $2|\Lambda|$ for the triangular
and ${4\over 3}|\Lambda|$ for the Kagom\'e lattice. This can be seen from
the following argument.
In general, $-n_{\rm max}$ is the ground state energy
of the {\em antiferromagnetic} Ising model under the restriction that the
magnetization is fixed, $\sum_{x\in \Lambda}\sigma_x=(2\rho-1)|\Lambda|$.
However, we do not need to deal with the
restriction. In both cases the (unrestricted) ground state is known to be
highly degenerate. Among the exponentially large number of ground state
configurations there are non-magnetized ones, corresponding to $\rho={1\over 2}$,
others with a concentration of up-spins $\rho={1\over 3}$, and between these
two limits $\rho$ can vary by steps of $1/|\Lambda|$. The rule is to
flip zero-energy spins one by one. The common
energy of all these configurations is easy to compute from the fact
that in each triangle there is precisely one unsatisfied bond. This fixes
the value of $n_{\rm max}$ as given above.

We also note that the optimal $x$ (which maximizes $B(v)$) is small: of
order 0.1 or smaller. We shall evoke this fact in the discussion of section 6.

\subsection{Lower bounds}

In their paper Dyson, Lieb and Simon (DLS) gave a lower bound on the
ground state energy of the spin-${1\over 2}$ XY-model (Dyson {\it et al} 1978,
Theorem C.1).
Because it corresponds to an upper bound on the norm of the Hamiltonian
(\ref{Ham}) in Fock space, it
is automatically valid for the hard-core boson gas ({\ref{Ham}) at any
density. It reads
\be\label{dls}
\e0\leq{1\over 4}|\Lambda|\left\{
\begin{array}{ll}
\sqrt{k(k+2)}&\mbox{if $k$ is even}\\
k+1&\mbox{if $k$ is odd\ .}
\end{array}   \right.
\ee
The above bound is to be compared with the trivial bound $\e0\leq n_{\rm max}$
obtained by writing Eq.~(\ref{eigen}) for an $X=X_0$ which maximizes $a_X$.
We then see that (\ref{dls}) is nontrivial for the $d$-dimensional hypercubic
lattice if $\rho>{1\over 4}\sqrt{1+{1\over d}}$, for the triangular lattice
if $\rho>{1\over 2\sqrt{3}}$, for the honeycomb lattice if $\rho>{1\over 3}$
and for the Kagom\'e lattice if $\rho>{1\over 4}\sqrt{3\over 2}$.

The bound (\ref{dls}) is the best at $\rho={1\over 2}$ and can be improved
for lower densities.

If $e_0$ is the ground state energy per site then
\be
|e_0|=\langle\Psi_0|A_x|\Psi_0\rangle
\equiv \langle\Psi_0|{1\over 2}\sum_{y\in\partial x}
(b_x^*b_y+b_y^*b_x)|\Psi_0\rangle
\ee
for any $x$ in $\Lambda$. $A_x$ preserves the number of bosons in
$\Lambda_x=\{x\}\cup\partial x$, therefore it commutes with
$P_{x,j}$, the orthogonal projection to the subspace ${\cal H}_{x,j}$
of states with $j$ particles in $\Lambda_x$, where $j=0,1,\ldots,k+1$.
We can write
\be
\langle\Psi_0|A_x|\Psi_0\rangle
=\sum_{j=0}^{k+1}{\langle\Psi_0|P_{x,j}A_xP_{x,j}|\Psi_0\rangle
\over\langle\Psi_0|P_{x,j}|\Psi_0\rangle}
\langle\Psi_0|P_{x,j}|\Psi_0\rangle\ .
\ee
Since $\langle\Psi_0|P_{x,j}\cdot P_{x,j}|\Psi_0\rangle
/\langle\Psi_0|P_{x,j}|\Psi_0\rangle$ is a normalized positive
linear functional,
\be\label{pap}
\langle\Psi_0|A_x|\Psi_0\rangle
\leq \sum_j\lambda_{\rm max}(j)\langle\Psi_0|P_{x,j}|\Psi_0\rangle\ ,
\ee
where $\lambda_{\rm max}(j)$ is the maximum eigenvalue of $A_x$ restricted
to ${\cal H}_{x,j}$. Now $\lambda_{\rm max}(j)$ was computed by DLS:
\be\label{lamj}
\lambda_{\rm max}(j)={1\over2}\sqrt{j(k+1-j)}\ .
\ee
The DLS-bound (\ref{dls}) corresponds to the maximum of (\ref{lamj}).
Let $h(t)$ be obtained by linear interpolation through the
points $(j,\lambda_{\rm max}(j))$:
\be\label{ht}
h(t)={1\over2}\left\{(\lfloor t\rfloor+1-t)
\sqrt{\lfloor t\rfloor(k+1-\lfloor t\rfloor)}
+(t-\lfloor t\rfloor)\sqrt{(\lfloor t\rfloor+1)(k-\lfloor t\rfloor)}\right\}\ .
\ee
Since $h(t)$ is a concave function, and the right
member of (\ref{pap}) can be written as $\int h(t)d\nu(t)$ with a probability
measure $\nu$ concentrated on the integers from $0$ to $k+1$, by Jensen's
inequality
\be
|e_0|\leq h\left(\int td\nu(t)\right)=h\left(
\sum_{j=0}^{k+1}j\langle\Psi_0|P_{x,j}|\Psi_0\rangle
\right)\ .
\ee
The expectation value in the argument of $h$ can be evaluated and yields
$(k+1)\rho$. Indeed, using the translation invariance of $\Psi_0$, after a
simple algebra we find
\be
\sum_{j=0}^{k+1}j\langle\Psi_0|P_{x,j}|\Psi_0\rangle
=\sum_X a_X^2{1\over \vol}\sum_{x\in\Lambda}|X\cap\Lambda_x|\ .
\ee
Let $\chi_X$ and $\chi_{\Lambda_x}$ denote the characteristic functions of
$X$ and $\Lambda_x$, respectively. Then
\be
|X\cap\Lambda_x|=\sum_{y\in\Lambda}\chi_X(y)\chi_{\Lambda_x}(y)
\ee
and therefore
\be
\sum_{x\in\Lambda}|X\cap\Lambda_x|=\sum_{y\in\Lambda}\chi_X(y)
\sum_{x\in\Lambda}\chi_{\Lambda_x}(y)=|X|\cdot|\Lambda_x|=N(k+1)\ .
\ee
Hence, we obtain
\be\label{hkr}
|e_0|\leq \min\{h((k+1)\rho),k\rho\}\leq{1\over2}(k+1)\sqrt{\rho(1-\rho)}
\ee
where the last member results from majorizing $h(t)$ by
${1\over2}\sqrt{t(k+1-t)}$.
Taking the minimum in Eq.~(\ref{hkr}) is not superfluous. For
$\rho<{1\over k+1}$ we have $h((k+1)\rho)={1\over 2}\sqrt{k}(k+1)\rho$.
Since $n_{\rm max}/|\Lambda|=k\rho$ is smaller
for all $k$, the trivial bound is better
for small densities. Actually the trivial bound provides the
right asymptotics at $\rho=0$
because $e_0=-k\rho+O(\rho^2)$ near $\rho=0$, as one can see by comparing
the trivial upper and lower bounds, $-M/\vol$ and $-n_{\rm max}/\vol$.

Numerical works on the spin-$1\over 2$ XY model in the square lattice yield
an approximate formula,
\be\label{MCFS}
e_0(\rho)=-1.09766+4.835(0.5-\rho)^2-1.99232(0.5-\rho)^4+0.85952(0.5-\rho)^6\ .
\ee
The first two terms come respectively from a Monte Carlo
(Zhang and Runge 1992)
and a finite-size scaling (Hamer {\it et al} 1999) study, to which
we have added the fourth and the sixth order terms to obtain
$e_0(0)=0$ and $e_0'(0)=-4$.
In Figure 1 we have plotted the
trivial upper bound $-4\rho(1-\rho)$,
the formula (\ref{MCFS}) and the lower bound
$-\min\{h((k+1)\rho),k\rho\}$. We have also shown numerical points for
a $5\times 5$ lattice from Table 1 of Hamer {\it et al} 1999. They nicely
follow the curve (\ref{MCFS}).

\begin{figure}
 \epsfxsize=9 truecm
 \centerline{\epsffile{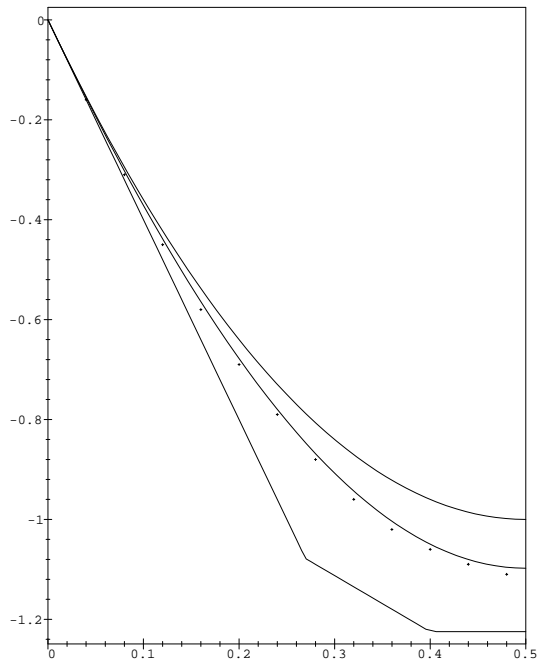}}
 \caption{Approximations of the ground state energy per site as a function of
the density in the square lattice.}
\label{fig1}
\end{figure}

Using the formula (\ref{MCFS}) we can compute the lower and upper bounds (\ref{lowup})
on $D_a^2/D^2$. Both of them increase monotonically as $\rho$ varies between
0 and 0.5: the lower bound from $\sim 0.66$ to 0.90234 and the upper bound
from 1 to 1.09766, their difference decreasing monotonically at the same time.

\section{Off-diagonal order and the final form of the ground state}

Off-diagonal long-range order in the ground state is characterized by
(cf. Appendix)
\be\label{odlro}
\omega_0\equiv\lim {1\over |\Lambda|^2}\sum_X a_X\sum_{x\in X}\sum_{y\notin X}
a_{X\cup\{y\}\setminus\{x\}}>0
\ee
where lim means the thermodynamic limit
($\vol$ and $N$ going to infinity and $N/\vol$ going to $\rho$).
In (\ref{odlro}) $\sum a_X^2=1$ is supposed.

Let us see, what would we obtain if $a_X$ did not fluctuate within
$\Omega_n$, i.e. if $a_X=\alpha_{\bndl}$. This and
(\ref{alen}) would imply another large-deviation principle in the approximate
form
\be\label{ldp^2}
\sum_{X\in\Omega_n}a_X^2 =\alpha_n^2 |\Omega_n|
\sim
\exp\left[-{(n-\lambda_2)^2\over 2D_{a^2}^2}\right]
\ee
with
\be\label{m2d2}
\lambda_2=\lambda_1+D_a^2{\lambda_1-M\over 2D^2-D_a^2}\qquad
D_{a^2}^2={D^2D_a^2\over 2D^2-D_a^2}\ .
\ee
The mean value $\lambda_2$ is larger than $\lambda_1$ because
\be\label{da2/d2}
{D_a^2\over D^2}\leq{\lambda_1\over M}<\min\left\{{1\over1-\rho}\ ,\
{k+1\over 2k\sqrt{\rho(1-\rho)}}\right\}<2\ ,
\ee
see Eq.~(\ref{hkr}).
As
\be\label{partial}
||\partial(X\cup\{y\}\setminus\{x\})|-\bndl|\leq 2k\ ,
\ee
with $a_{X\cup\{y\}\setminus\{x\}}\geq\alpha_{\bndl-2k}$ we would
find
\be\label{nofluc}
\omega_0\geq\rho(1-\rho)\lim e^{-{2k(\lambda_1-M)\over 2D^2-D_a^2}}>0\ .
\ee
The argument above, together with the conclusion (\ref{nofluc}) may be
right in high enough dimensions
but it is known to be incorrect in one dimension.
Fluctuations have to be taken into account.

A way to prove (\ref{odlro}) would be to show
\be\label{ageqa}
a_{X\cup\{y\}\setminus\{x\}}\geq c a_X
\ee
for all $X$, $x$ and $y$ with a $c>0$ independent of $\Lambda$ and $N$.
If (\ref{ageqa}) did hold true, we would find
$\omega_0\geq c\rho(1-\rho)>0$. In the complete graph (\ref{ageqa})
is verified with $c=1$ and equality sign.
However, we know already from the considerations following Eq.~(\ref{eigen})
that in finite dimensional lattices (\ref{ageqa}) cannot hold indeed for all
$X$: e.g., in the case when $\bndl\propto N^{1-{1\over d}}$, (\ref{ageqa})
can be satisfied only with $c\propto N^{-{1\over d}}$. Also, we should not
be able to prove (\ref{ageqa}) in one dimension. Below we present an
argument which takes into account
fluctuations and distinguishes between one, two and higher dimensions.

First, we note that we need (\ref{ageqa}) only for
$X$ in a subset $S(\Lambda,N)$ such that the sum of $a_X^2$ over $S$
is non-vanishing in the thermodynamic limit.
Let us introduce a function $R(X)$ by setting $a_X=\alpha_{\bndl}e^{R(X)}$.
Then
\be\label{RX}
\sum_{X\in\Omega_n}e^{R(X)}=|\Omega_n|\leq\sum_{X\in\Omega_n}e^{2R(X)}\leq |\Omega_n|^2\ ,
\ee
so we can define an $\epsilon_n$ such that $0\leq\epsilon_n\leq 1$ and
\be\label{epsfirst}
\sum_{X\in\Omega_n}e^{2R(X)}=|\Omega_n|^{1+\epsilon_n}\ .
\ee
Using this and Eq.~(\ref{alen}) we find
\be\label{aepsn}
\sum_{X\in\Omega_n}a_X^2=\alpha_n^2 |\Omega_n|^{1+\epsilon_n}
\sim\exp\left[-{(n-M_n)^2\over 2D_n^2}\right]
\ee
with
\be
M_n=\lambda_1+{(1-\epsilon_n)D_a^2\over 2D^2-(1-\epsilon_n)D_a^2}(\lambda_1-M)\qquad
D_n^2={D^2D_a^2\over 2D^2-(1-\epsilon_n)D_a^2}\ .
\ee
Because of the possible $n$-dependence of $M_n$ and $D_n$, Eq.~(\ref{aepsn})
may not describe a large deviation principle. However,
$M_n$ and $D_n$ satisfy the inequalities
$\lambda_1\leq M_n\leq \lambda_2$ and
$D_a/\sqrt{2}\leq D_n\leq D_{a^2}$ and, hence, for any $\varepsilon>0$
\be\label{epslast}
\lim
\sum_{(1-\varepsilon)\lambda_1\leq\bndl\leq(1+\varepsilon)\lambda_2}a_X^2=1\ .
\ee
For $S$ we can take the set of summation with any $\varepsilon\geq 0$.

Next, we observe that by using Eqs.~(\ref{nonbipar}) and (\ref{bipar})
we can write
\be\label{ratio}
{a_X\over a_Y}
=\lim_{m\rightarrow\infty}{W_m(X)\over W_m(Y)}\quad{\rm or}\quad
\lim_{m\rightarrow\infty}{\lambda_1W_m(X)+W_{m+1}(X)
\over \lambda_1W_m(Y)+W_{m+1}(Y)}
\ee
for non-bipartite or bipartite lattices, respectively.
The convergence in (\ref{ratio}) is usually slow.
However, we expect that if $\bndl$ and $|\partial Y|$ do not differ too much then,
at least up to the order of magnitude, $a_X/a_Y$ can be approximated
by the ratio of the number of walks whose length is the
distance between $X$ and $Y$. Thus,
we conjecture that for $\bndl=c_1 N$ and $||\partial Y|-\bndl|\leq c_2$
\be\label{conj}
\left|\ln{a_X\over a_Y}-\ln {W_{d_G(X,Y)}(X)\over W_{d_G(X,Y)}(Y)}\right|
\leq c_3
\ee
where $c_3$ may depend on the constants
$c_1$ and $c_2$ but not on the size of the system.
In what follows, we examine the consequences of this hypothesis.

If the aspect ratios of $\Lambda$ are kept bounded,
its diameter $L$ is of order
$|\Lambda|^{1\over d}$ and the diameter of $G$ is of order
$L^{d+1}$. Over such a distance (\ref{conj}) allows a gigantic change of $a_X$.
However, the distance to be considered
for (\ref{ageqa}) is much smaller. If the symmetric difference of $X$
and $Y$ is $\{x,y\}$, it is easily seen that $d_G(X,Y)=d(x,y)\leq L$.
Let $\bndl=n\propto N$.
Above one dimension we apply the inequalities (\ref{Wnineq})
to obtain
\be\label{W/W}
{W_{d_G(X,Y)}(X)\over W_{d_G(X,Y)}(Y)}\leq
\left({1+2(k-1)L/n\over 1-[2(k-1)L+2]/n}\right)^{L}\ .
\ee
In the thermodynamic limit the upper bound remains finite
in two dimensions and converges to 1 in higher dimensions,
so we get the necessary estimate (\ref{ageqa}).
In one dimension there exists no convergent upper bound.
Thus, we find ODLRO in the ground state in two and higher dimensions
but not in one dimension.

The difference between
2 and $2+\varepsilon$ dimensions in the convergence of the right-hand side of
Eq.~(\ref{W/W})
indicates that $d=2$ is the critical dimension above which
the deviations of $a_X$ from the average $\alpha_{\bndl}$
are irrelevant. This remark can be made quantitative if we observe
an interesting consequence of the hypothesis (\ref{conj}) on $R(X)$.
If $\bndl\propto N$ then inside a sphere of radius $\propto \sqrt{N}$, centred
at $X$, for all $Y$ such that $||\partial Y|-\bndl|$ is bounded by some constant
of the order of unity $|R(Y)-R(X)|$ also remains below another constant of order~1.
Comparing the scale $\sim\sqrt{N}$ with the diameter $\sim N^{1+1/d}$ of $G$
we conclude that
\be\label{maxrx}
|R(X)|\leq\max R(Y)-\min R(Y)<{\rm const}\times N^{{1\over 2}+{1\over d}}\ .
\ee
The maximum and the minimum are taken over $\Omega_{\bndl}$.
The first inequality holds because for any $n$ the average of $R(X)$ over
$\Omega_n$ is negative, but $R(X)$ cannot be negative for all $X$.

According to Equation~(\ref{maxrx}),
\be\label{afinal}
a_X=\exp[F(\bndl)+R(X)]=
\exp\left[F(\bndl)+O\left(N^{{1\over2}+{1\over d}}\right)\right]\ .
\ee
Therefore, in one dimension $R(X)$
can dominate $F(\bndl)$, in two dimensions it can vary on the same
scale $\sim N$,
and above two dimensions the variation of $R(X)$ is
negligible compared with that of $F(\bndl)$. Apart from an additive
constant necessary for normalization, the approximate form of
$F(n)=\ln\alpha_n$ can be read off from Eq.~(\ref{alen}).

The upper bound in the right-hand member of Eq.~(\ref{maxrx}) could, of course,
overestimate the order of magnitude of the maximum of $|R(X)|$. However,
because it seems to correctly distinguish between one, two and higher
dimensions, it is probably sharp.
This implies also that on the scale $\sim\sqrt{N}$ the function $R(X)$ can
be considered as describing a ballistic motion over $G$ rather than a
random fluctuation about a negative average,
which would lead to a smaller bound. Accepting
ballisticity on smaller scales as well, one can
find the order of magnitude of local variations
of $R(X)$. We distinguish between two cases.

(a) If $\Lambda$ is bipartite with sublattices $V_1$, $V_2$ then $G$ is also
bipartite with sublattices
$\Gamma_1=\{X:|X\cap V_1|\ \mbox{is even}\}$ and
$\Gamma_2=\{X:|X\cap V_1|\ \mbox{is odd}\}$. Along each path in $G$ the subsequent
terms belong to alternating sublattices. If $\partial\,^2 X$ denotes the set
of second neighbours of $X$ then
\be\label{cons1}
|R(X)-R(Y)|<{\rm const}\times N^{-1/2}\quad\mbox{if $Y\in\partial\,^2X$ and $|\partial Y|=\bndl$}\ .
\ee

(b) If $\Lambda$ is not bipartite, neither is $G$, and no systematic
compensation of terms of different signs is possible. In this case
\be\label{cons2}
|R(X)-R(Y)|<{\rm const}\times N^{-1/2}\quad\mbox{if $Y\in\bnd$ and $|\partial Y|=\bndl$}\ .
\ee
The result on ODLRO could be obtained from (\ref{cons1}), (\ref{cons2}) as well.

As the two-dimensional case is marginal, some subtle logarithmic corrections,
that we do not see, may modify our conclusions.

Above two dimensions we also find, as a further consequence
of Eq.~(\ref{maxrx}), that $\epsilon_n$
identically vanishes in the thermodynamic limit. Thus, $M_n/\lambda_2\rightarrow 1$
and $D_n/D_{a^2}\rightarrow 1$, Eq.~(\ref{aepsn}) asymptotically coincides
with Eq.~(\ref{ldp^2}), Eq.~(\ref{nofluc}) provides a valid lower bound
on the order parameter, and Eq.~(\ref{epslast}) can be replaced by
\be\label{eps22}
\lim
\sum_{(1-\varepsilon)\lambda_2\leq\bndl\leq(1+\varepsilon)\lambda_2}a_X^2=1
\ee
for any $\varepsilon>0$. This property makes $a_X^2$ similar to a thermal
Gibbs state for a classical lattice gas. The classical Hamiltonian
(with $1/k_B T$ incorporated) is $\Heff(X)=-2F(\bndl)-2R(X)$, and there is
asymptotic equivalence between the canonical distribution and a microcanonical one
concentrated on $X$'s with $\Heff(X)$ in an $o(N)$ neighbourhood of the
canonical expectation value of $\Heff$. This holds in spite of $\Heff$
containing long-range interactions. Indeed, $-F$ is a sum of
nearest neighbour repulsive two-body interactions, $f_x$, and, if $D_a\neq D$,
properly normalized
long-range four-body interactions, $f_xf_y$, see Eqs.~(\ref{fx}),
(\ref{DeltaX}) and (\ref{alen}), the whole yielding a global repulsion
which tends to maximize $\bndl$. Now $R$ may also include up to $N$ body
interactions but because $R(X)=o(N)$ {\em and because $a_X^2$ corresponds
to a high-temperature Gibbs state}, it can be neglected: The fact that
$R$ may partly split the ground state degeneracy of $-F$ is irrelevant at high
temperatures.

We have no rigorous proof that the limiting effective Gibbs state is a
pure (in particular, a high-tem\-per\-ature)
state. For this we should show that for all densities and all relevant $n$,
$2F(n)/n<K_c$, the critical coupling
of the Ising model on the same lattice. Even if we knew $\lambda_1$ exactly,
our bounds (\ref{lowup}) on $D_a/D$ are not good enough to obtain this
information. For example, in the case of the square lattice we can use
the numerical fit (\ref{MCFS}). If we assume that $D_a\geq D$, we get
$2F(n)/n\leq 2(\lambda_1-M)/D^2$ where the upper bound increases with $\rho$
and its maximum at $\rho=0.5$ is $0.3906<K_c=0.4407$. On the other hand,
if we use the lower bound for $D_a$, at half filling we find for $n\geq\lambda_1$
that $2F(n)/n\leq 0.628>K_c$. An indication to
the high-temperature character can be found in the variational estimates,
cf. the final remark
of section 5.1. In the light of Eq.~(\ref{afinal}), trial functions of the
form (\ref{trial}) acquire a particular importance: For $d>2$ one could find
with them the exact ground state energy per site, and the optimal exponential
ansatz may not be far from the true ground state. The variable $x$ appearing in
the formula (\ref{av*av}) corresponds to ${1\over 2}\beta J$ of the Ising model.
The estimates we have done predict a value well below the critical one in
any dimension: $\beta J< 0.2$ for all densities at $k=3$, and decreases with
increasing $k$, reaching 0 in the complete graph.

Now the fact that $a_X^2$ is equivalent to a high-temperature Gibbs state means that there
is no classical (diagonal) order in the ground state, coexisting with
the purely quantum-mechanical (off-diagonal) one.
Diagonal order is characterized by an order
parameter operator $O$ which is diagonal in the basis (\ref{basis})
and for which the ground state expectation value
$\vol^{-2}\langle\Psi_0|O^*O|\Psi_0\rangle=\vol^{-2}\sum a_X^2 |O(X)|^2$ has a
non-vanishing thermodynamic limit. A typical example is
$O_q=\sum_x e^{i(q,x)}n_x$,
associated with a periodic order with a wave vector~$q$. Due to the
correspondence with a Gibbs state, diagonal order in the ground
state of the Bose gas is equivalent to low-temperature order in
the associated classical lattice gas.
The ground state of $-F$ is exponentially degenerate for $\rho<1/2$.
However, it is only two-fold degenerate on bipartite lattices at $\rho=1/2$,
and the degeneracy is not split by the translation invariant $R(X)$.
Apparently, there is no
qualitative argument against a crystalline order at half filling
in the ground state of the $d>2$ dimensional purely hard-core Bose gas.
The nearest neighbour repulsion in $-F$ is simply not strong enough,
reflecting the fact that a purely on-site repulsion in $H_0$,
even though infinite, cannot induce such an order.

\section{Extensions and concluding remarks}

The nonuniformity of the ground state we have found in a
purely hard-core gas seems to be a general property of
finite-range Hamiltonians. The extension to more complicated
hard-core Hamiltonians is immediate. Let
\bea\label{Hgen}
H&=&-\sum_{\langle xy\rangle}(b_x^*b_y+b_y^*b_x)+\sum V_x n_x
-J\sum_{\langle xy\rangle}\left(n_x-{1\over 2}\right)\left(n_y-{1\over 2}\right)
+K(\{n_x\})\nonumber\\
&\equiv& H_0+V+{J\over 4}(2\Delta-|E\Lambda|)+K
\eea
where $\Delta$ is the diagonal
part of $H_0^2$, cf. Eq.~(\ref{diag}), and $K$ may contain further
interactions of a bounded range.
Because the terms additional to $H_0$ are diagonal in the
basis (\ref{basis}), the Perron-Frobenius theorem still applies and
the ground state wave vector $a_X>0$. If $A$ is the matrix of $-H$ then
keeping the definition (\ref{W}) we still have (\ref{EaWn}), and the
inequalities of section 2 also hold with suitable modifications.
In particular, in (\ref{parxy}) and (\ref{w1s-w2}) $2(k-1)$ is to be replaced
by $\kappa=|2-J|(k-1)+\delta V+\delta K$ where
$\delta V=\max_{\langle xy\rangle}|V_x-V_y|$ and
$\delta K=\max_{X,Y\in\bnd}|K(Y)-K(X)|$
are of the order of unity, and in (\ref{Da})-(\ref{Ch2}) we have
$\kappa\langle\bndl\rangle_a$ instead of $2(k-1)\lambda_1$. Now
\be
W_1(X)=A_{XX}+\bndl=\left(1-{J\over 2}\right)\bndl-V(X)-K(X)
+{J\over 4}|E\Lambda|
\ee
may not be positive and $\langle W_1\rangle_a=\lambda_1=-E_0$ may not be of order $N$.
Therefore, there may not be a large deviation principle for the pair $(W_1,a)$.
However, if we replace $H$ by $H^c=H+cI$ where $I$ is the identity operator, the
ground state vector $a$ will not change. The corresponding $W_1$ is $W_1^c(X)=W_1(X)-c$,
and $D_a^2(W_1^c)$ is also independent of $c$. Let us choose $c=\langle A_{XX}\rangle_a$,
then
\be
\lambda_1+c\equiv\lambda_1^c=\langle W_1^c\rangle_a=\langle\bndl\rangle_a
\ee
and, thus, we have a large deviation principle for $(W_1^c,a)$ in the same form as
for the purely hard-core interaction.

What really counts for the non-uniformity of $a$ is not the large deviation
principle for $(W_1,a)$ but the variation of $W_1(X)$ over the set of $N$-point
configurations. The ground state is nonuniform if $[\max W_1(X)-\min W_1(X)]/\sqrt{N}$
goes to infinity with $N$. As an example, let us consider the nearest neighbour
anisotropic Heisenberg model ($V=K=0$). At $J=2$, when it is isotropic,
for any fixed $N$ we have
$W_1(X)\equiv |E\Lambda|/2$ and hence $\lambda_1=|E\Lambda|/2$
and $a_X$ is constant.
Away from the isotropy point $\Delta W_1$ is of order $N$
and $a$ is nonuniform.

If $\Omega_n=\{X: W_1(X)=w_n\}$,
where $w_1<w_2<\ldots$ are the possible values of $W_1$, one can write
down an approximate formula similar to Eq.~(\ref{qan}), namely
$q(n)\sim\exp[-(w_n-\overline{W_1})^2/2D^2(W_1)](w_{n+1}-w_n)$.
Interpreting $\alpha_n$ as the average of $a_X$ over the newly defined
$\Omega_n$, we have also the analogue of Eq.~(\ref{alen}).

As earlier, we can write
$a_X=\exp[F(W_1(X))+R(X)]$, and Eq.~(\ref{RX}) remains valid.
The approximate form of $F$ is
\be\label{F}
F(W_1(X))\approx
{(W_1(X)-\overline{W_1})^2\over 2D(W_1)^2}-
{(W_1(X)+E_0)^2\over 2D_a(W_1)^2}+\,{\rm const}
\ .
\ee

One can replace the infinite on-site repulsion by a finite one
provided that stability is assured, $H>-\,{\rm const}\times N$.
The dimension of the $N$-particle subspace will increase
to $\vol+N-1\choose N$. A
configuration $X$ becomes a list of $N$ elements containing vertices
of $\Lambda$ with possible repetitions that we can write as
$X=\{x^{n_x}\}_{x\in\Lambda}$. The neighbourhood relation
remains the same: $Y\in\bnd$ if $Y$ can be obtained
from $X$ by moving a single particle along an edge of $\Lambda$, say,
from $x$ into $y$. The corresponding matrix element of $-H$ is then
$A_{XY}=\sqrt{n_x(X)n_y(Y)}$. Now $\sum_{Y\neq X}A_{XY}$
can vary between $k\sqrt{N}$ and $\sqrt{2}kN$ when $\bndl$ varies
between $k$ and $kN$. However, $A_{XX}$ can be of order $N^2$ and
its change between neighbours of order $N$. Therefore, we miss
a pointwise bound like (\ref{parxy}) and cannot repeat the argument of
section 2. Nevertheless, $-W_1(X)=\sum_Y\langle Y|H|X\rangle$ is still the
energy of a classical lattice
gas in which the equivalence of the canonical and microcanonical
ensembles implies the analogue of Eq.~(\ref{qan}) and, via the
argument of section 4, the nonuniformity of the ground state.

The real novelty introduced by the additional interactions is the possible
coexistence of classical long-range order with the off-diagonal one. A
thorough discussion of this question, concluding negatively, can be found
already in Penrose and Onsager's 1956 paper. Nevertheless, a controversy
has remained until recently, when reliable quantum Monte Carlo computations
on square lattice models have shown such a coexistence (Batrouni {\it et al} 1995,
Scalettar {\it et al} 1995).
We hope that the ideas developed in this paper
can contribute in the future to further elucidate this interesting problem.

\vspace{2mm}

This work was supported by the Hungarian Scientific Research Fund (OTKA)
under Grant No. T 30543.

\section*{Appendix. A digression on ODLRO}

Following the standard
definition (Penrose and Onsager 1956, PO), we give below the expression
of the ODLRO parameter in terms of $(a_X)$, valid for any Bose gas whose
Hamiltonian contains a hard-core repulsion and conserves the number of particles.
The density operator in the ground state $\Psi_0$ (supposed
to be normalized) is
the orthogonal projection $|\Psi_0\rangle\langle\Psi_0|$. The density
matrix in the basis (\ref{basis}) is therefore $(a_Xa_Y)$.
From here the one-particle reduced density matrix $\sigma=(\sigma_{xy})$
is obtained by taking a partial trace over $N-1$ particle positions,
\begin{equation}\label{onepart}
\sigma_{xy}=\sum_{X'}a_{X'\cup\{x\}}a_{X'\cup\{y\}}\ .
\end{equation}
In the above sum $X'$ runs over the $N-1$-point subsets of $\Lambda$
not containing $x$ and $y$. The matrix $\sigma$ is real symmetric, positive
semidefinite (in fact, positive definite for $N>2$), its trace is $N$ and
all its elements are positive.
A possible choice for the order parameter, according to PO, is
\begin{equation}\label{omega}
\omega=|\Lambda|^{-2}\sum_{x,y\in\Lambda}\sigma_{xy}\ .
\end{equation}
Another definition of PO for the order parameter is $1/|\Lambda|$
times the largest eigenvalue of $\sigma$.
If $\Lambda$ is a shift-invariant set (which
supposes periodic boundary conditions), the ground state is translation
invariant and $\sigma_{xy}=\sigma_{x-y}$.
Applying the Perron-Frobenius theorem to $\sigma$ we find that its largest
eigenvalue
is $\sum_y\sigma_{x-y}=|\Lambda|\omega$,
so that the above two definitions coincide.
The order parameter
$\omega$ also has an interpretation as the density of the
Bose-Einstein condensate. This holds because the operator
associated with the density of the condensate is
\begin{equation}\label{condop}
|\Lambda|^{-1}b^*_{k=0}b_{k=0}=|\Lambda|^{-2}\sum_{x,y\in\Lambda}b_x^*b_y
\end{equation}
and because
\begin{equation}\label{sigmaxy}
\langle\Psi_0|b_x^*b_y|\Psi_0\rangle=\sigma_{xy}\ .
\end{equation}
Inserting (\ref{onepart}) into (\ref{omega}) we obtain
\be\label{om'}
\omega=\omega'+{N\over|\Lambda|^2}\qquad
\omega'={1\over|\Lambda|^2}\sum_X a_X \sum_{x\in X}\sum_{y\notin X}
a_{X\cup\{y\}\setminus\{x\}}\ .
\ee
The maximum of $\omega'$ under the condition of normalization is $\rho(1-\rho)$,
attained with $a_X$ constant. It is realized by $H_0$ on complete graphs and
by the ferromagnetic Heisenberg model on arbitrary graphs.
Our $\omega'$ is one fourth of the order parameter usually used in the
spin-$1\over 2$ XY model, cf. Fujiki and Betts 1986, Hamer {\it et al} 1999.
There is ODLRO for a density $\rho$ if $\omega$ or, equivalently,
$\omega'$ has a
nonvanishing limit $\omega_0>0$ as $N$ and $|\Lambda|$ go to infinity while
$N/|\Lambda|$ goes to $\rho$. When this occurs, conservation
of the particle number
is spontaneously broken and there appears (at least) a
one-parameter continuum of ground states in infinite volume. The order
parameter operator, $\sum b_x^*$, is nondiagonal in the natural basis
(\ref{basis}) (whence the name ODLRO).
The ground states
$\gamma_\alpha$, $0\leq\alpha<2\pi$, are symmetry-related. For any local
operator $B$
\be
\gamma_\alpha(B)=\gamma_0\left(\prod_x e^{-i\alpha n_x}B
\prod_x e^{i\alpha n_x}\right)\ .
\ee
In particular,
\be
\gamma_\alpha(b_x^*)=e^{-i\alpha}\gamma_0(b_x^*)\neq 0\ .
\ee

\end{document}